\documentstyle[12pt]{article}
\def\G{\Gamma}

\def\P{\Phi}
\def\p{\phi}

\def\da{\dag}
\def\d{\delta}
\def\a{\alpha}

\def\ba{\begin{array}}
\def\ea{\end{array}}
\def\hs{\hspace}

\def\ot{\otimes}

\def\o{\Omega}
\def\ll{\label}

\def\ti{\times}
\def\be{\begin{equation}}
\def\ee{\end{equation}}
\def\bea{\begin{eqnarray}}
\def\eea{\end{eqnarray}}
\def\nn{\nonumber}
\def\f{\frac}

\begin{document}

\parskip 5pt plus 1pc
\parindent=20pt
\begin{flushright}
AS-ITP-94-28\\
August, 1994
\end{flushright}
\vspace{5ex}

\centerline{\Large \bf Gauge theory on $Z_2 \times Z_2 \ti Z_2 $ Discrete
Group}
\centerline{\Large\bf  and a Spontaneous $CP$ Violation Toy Model }

\vspace{10ex}
\centerline{\large \bf Bin Chen$^a$~~, \hs{2ex} Hao-Gang Ding$^b$~~ and
Ke Wu$^a$}
\centerline{\normalsize \bf a. Institute of Theoretical Physics
Academia Sinica}
\centerline{\normalsize \bf P.O.Box 2735,Beijing 100080,China
\footnote{The project is supported by National Natural Science Foundation
of China and "Climbing Up" Foundation of China}}
\centerline{\normalsize \bf  b. Department of Physics, Peking University,
Beijing 100871 China;\footnote{\sf Mailing address}\\}

\vspace{20ex}
\centerline{\large\bf Abstract}
\vspace{3ex}
\begin{minipage}{5in}
\it In the spirit of Non-commutative differential
calculus on discrete group, we construct a toy model of spontaneous $CP$
violation (SCPV). Our model is different from the well-known
Weinberg-Branco model although it involves three Higgs doublets and
preserve neutral flavor current conservation (NFC)
after using the $Z_2 \ti Z_2 \ti Z_2$ discrete symmetry
and imposing some constraints on Yukawa couplings.
\end{minipage}
\vspace{4ex}

\newpage

\section{Introduction}

\hskip20pt Ever since the discovery of $CP$ violation, many mechanisms have
been
put forward to explain it. Among them, Kobayashi-Maskawa(K-M) model \cite{K-M}
is the most famous one and has been accepted as the standard model.
However, because all
experimental data indicate that the origin of $CP$ violation is like to be of
superweak type, K-M model is not a natural one for its milliweak character
despite it behave like a superweak model.

Among other mechanisms, the Weinberg-Branco model \cite{Wein}
of SCPV has attracted much
attention for its theoretical value. There are three Higgs doublets in their
model, and in order to achieve so-called "natural flavor conservation"
a discrete symmetry is imposed to eliminate the flavor changing neutral
current. The dominant CP violation mechanisms is mediated by the charged Higgs
bosons. As a result the
model is milliweak in character and predicts large $CP$ violation
in many experiments
other than the neutral kaon system. Therefore it has been given a very tight
constraints from the experiments. Very rescently, Y-L Wu propose a $SU(2)_L \ti
U(1)_Y$ gauge theory with two Higgs doublets which has both the SCPV and the
NFC
\cite{Wu,Wu2}.
He pointed that the Glashow-Weinberg criterion for NFC is to be only sufficient
but not necessary. After imposing some constraints on Yukawa coupling matrix,
his model allow each Higgs doublet couple to all the fermions so that the
lagrangian does not possess any additional discrete symmetries.

None of the existing $CP$ violation mechanisms is good enough to explain all
experimental data, $CP$ violation remains a mystery. We think that this fact is
related to our lack of knowledge on the origin of Higgs particle. Fortunately
with the help of non-commutative geometry some improvements have been made to
understand the origin of Higgs particle$[1\sim 5]$.
The Higgs field can be viewed as the gauge
field with respect to the discrete $Z_2$ group\cite{Sit,Ding}.
 Due to the new discrete
symmetry, the potential can be calculated within less freedom. This allow us to
understand spontaneous violation better.

In this paper we construct a toy model to explore the possible relation between
the SCPV and discrete symmetry in the spirit of non-commutative geometry.
In our model, we take the discrete group to be $Z_2 \ti Z_2 \ti Z_2$,
and following the formalism in \cite{Ding}, we get the lagrangian that a
dmits SCPV. After using
the result of Y-L Wu, we can preserve NFC in our model but allow each Higgs
doublet couple to all flavor of fermions.

This paper is scheduled as follows: In section 2, we briefly give out the
result
of gauge theory on discrete group $Z_2 \ti Z_2 \ti Z_2$.
In section 3, we construct our toy model.
At last section, we
end with some conclusions and remarks.

\section{The Gauge theory on the Discrete Group $Z_2 \ti Z_2 \ti Z_2$}

\subsection{ Differential Calculus on Discrete Groups $G$}
\hskip20pt Let $G$ be a discrete group of size $N_G$, its elements are
$\{e, g_1, g_2, \cdots , g_{N_G-1}\}$, and $\cal A$
the algebra of
the all complex valued functions on $G$ .
The basis $\partial_i, ~(i=1, \cdots , N_G-1) $
of the left invariant vector space $\cal F$ on $ \cal A$ are defined as
\be
\partial_if=f-R_if,~~\forall f\in \cal A, \ee
where
\be
(R_if)(g)=f(g\cdot g_i). \ee
Obviously $\partial_i$ is nothing but the difference operator acting
on $\cal A$, and satisfies
\be
\partial_i \partial_j=\displaystyle\sum_k C_{ij}^k
\partial_k,~~~~ C_{ij}^k=\delta _i^k + \delta _j^k -
\delta _{i\cdot j }^k\ee
where  $i, j, \cdots, (i\cdot j)$ denote  $g_i, g_j, \cdots,
(g_i\cdot g_j)$ respectively.
Let $\Omega^1$ be the dual space of $\cal F$, whose
 basis $\chi^i$  are one-forms, satisfy
\be
 \chi^i(\partial_j)=\delta^i_j. \ee

The exterior derivative on $\cal A$ is given by
 \be
df = \sum_{i=1}^{N_G-1}\partial_i f \chi^i.
\ee
The nilpotency of d and the graded Leibniz rule
\be
\begin{array}{cl}
(i) ~~&d^2 =0,\\[3mm]
(ii)~~&d(fg)= df\cdot g + (-1)^{deg f}f\cdot dg,~~~~~\forall f, g \in \Omega^*,
\\[4mm]
\end{array}
\ee
could be obtained provided that $\chi^i$ satisfy the following two
conditions\cite{Sit}
\be
\begin{array}{cl}
\chi^if=&(R_if)\chi^i,~~\forall f\in \cal A,\\[3mm]
d\chi^i=&-\sum_{j,k}C_{jk}^i\chi^j\otimes \chi^k.
\end{array}
\ee

The involution operator * on the differential algebra is
well defined if it
agrees with the complex conjugation on ${\cal A}$, takes the assumption
that $(\chi^g)^*=-\chi^{g^{-1}}$,
and (graded) commutes with d, i.e. $d(\omega^*)=(-1)^{deg\omega}(d\omega)^*$.
The Haar integral, which remains invariant under group action, is introduced
as a complex valued linear functional on $\cal A$ as,
\be
\int_G f=\frac 1 {N_G}\sum_{g\in G}f(g).
\ee

\subsection{ Gauge Theory on Discrete Group $G$}
Let us now construct the generalized gauge theory on finite groups
using the above differential calculus.
Like the usual gauge theory, the
 $d+\phi$ is gauge covariant which requires the transformation of
 gauge field one form $\phi$ as,
\be
\phi \rightarrow H\phi H^{-1}+HdH^{-1}. \ee
If we write $\phi =\displaystyle{\sum_g}\phi_g \chi^g$,
the coefficients $\phi_g$ transform as
\be
 \phi_g \rightarrow H\phi_g(R_gH^{-1})+H\partial_gH^{-1}.
\ee
It is convenient to introduce a new field $\Phi_g=1-\phi_g$, then
(2.10) is equivalent to
\be
\Phi_g \rightarrow H\Phi_g(R_gH^{-1}). \ee
The extended anti-hermitian condition $\phi^* = -\phi$ results in the
following relations on its coefficients $\phi_g$ as well as $\Phi_g$,
\be
\phi_g^{\dag} = R_g(\phi_{g^{-1}}), ~~~~\Phi_g^{\dag}=R_g(\Phi_{g^{-1}})
\ee
which will be very useful in following discussions.

It can be easily shown that the curvature two form
{}~~$ F=d\phi+\phi \otimes \phi $~~
is gauge covariant and can be written in terms of its  coefficients
\be
 F=\displaystyle{\sum_{g,h}}F_{gh}\chi^g\otimes\chi^h \ee
\be
F_{gh}=\Phi_gR_g(\Phi_h)-\Phi_{h\cdot g}.\ee

In order to construct the Lagrangian of this gauge theory on
discrete groups
we need to introduce a metric on the forms. Let us first define the
metric $\eta$ as a bilinear form on the bimodule $\Omega^{1}$ valued in the
algebra ${\cal A}$,
\be
\eta : ~~\Omega^1 \otimes \Omega^1 \rightarrow \cal A \ee
\be
 <\chi^g, \chi^h>=\eta^{gh}.
 \ee
The gauge invariance requires that $\eta^{gh} \sim
\delta^{gh^{-1}}$\cite{Ding}. However, the metric on the two forms becomes
more complicated, if only the condition of gauge invariance is required.
\be
<\chi^g \otimes \chi^h, \chi^p \otimes \chi^q >=
\alpha \eta^{gh}\eta^{pq} + \beta \eta^{gq}\eta^{hp} +
\gamma \eta^{gp}\eta^{hq},\ee
where the term proportional to $\gamma$ is only appeared when $G$ is
commutative\cite{Sit}. Then the most general Yang- Mills action is given,
\be
\ba{cl}
 {\cal L}&=\displaystyle-\int_G <F,\overline F> \\[3mm]
&=\displaystyle-\int_G \sum_{g,h,p,q} Tr(F_{gh}F^{\dag}_{pq})
<\chi^g \otimes \chi^h, \chi^{g^{-1}}\otimes
\chi^{p^{-1}}>\\[3mm]
&=\displaystyle-\int_G \sum_{g,h,p,q} Tr(F_{gh}F^{\dag}_{pq})
(\alpha \eta^{gh}\eta^{q^{-1}p^{-1}}
+ \beta \eta^{gp^{-1}}\eta^{hq^{-1}} +
\gamma \eta^{gq^{-1}}\eta^{hp^{-1}}),
\ea \ee
where we have used the involution relations
\be
{\overline F}=(\chi^q)^* \otimes(\chi^p)^* F^{\dag}_{pq},~~~~~
(\chi^g)^*=-\chi^{g^{-1}}.\ee

\subsection{ Gauge Theory on $Z_2\times Z_2\times Z_2$}
\hskip20pt We apply the general discussion in section 2.2 to the case of
$G=Z_2\times Z_2\times Z_2$. There are 8 elements in $G$, which could be
denoted as $\{ g_0, g_1, g_2, g_3, g_4, g_5, g_6, g_7\}$, or written in
terms of the elements $\{ e,r\}$ in $Z_2$,
\be\ba{cl}
g_0&=\{ e,e,e\}, \hspace{5ex} g_1=\{ r,e,e\}\\[3mm]
g_2&=\{ e,r,e\}, \hspace{5ex} g_3=\{ e,e,r\}\\[3mm]
g_4&=\{ r,r,e\}, \hspace{5ex} g_5=\{ r,e,r\}\\[3mm]
g_6&=\{ e,r,r\}, \hspace{5ex} g_7=\{ r,r,r\}\\[3mm]
\ea\ee
The multiplication rule of $Z_2\times Z_2\times Z_2$ is given in the
following table,\\
\vspace{1ex}
\begin{center}
\begin{tabular}{||c|c|c|c|c|c|c|c|c||}
\hline
     &$g_0$&$g_1$&$g_2$&$g_3$&$g_4$&$g_5$&$g_6$&$g_7$\\
\hline
$g_0$&$g_0$&$g_1$&$g_2$&$g_3$&$g_4$&$g_5$&$g_6$&$g_7$\\
\hline
$g_1$&$g_1$&$g_0$&$g_4$&$g_5$&$g_2$&$g_3$&$g_7$&$g_6$\\
\hline
$g_2$&$g_2$&$g_4$&$g_0$&$g_6$&$g_1$&$g_7$&$g_3$&$g_5$\\
\hline
$g_3$&$g_3$&$g_5$&$g_6$&$g_0$&$g_7$&$g_1$&$g_2$&$g_4$\\
\hline
$g_4$&$g_4$&$g_2$&$g_1$&$g_7$&$g_0$&$g_6$&$g_5$&$g_3$\\
\hline
$g_5$&$g_5$&$g_3$&$g_7$&$g_1$&$g_6$&$g_0$&$g_4$&$g_2$\\
\hline
$g_6$&$g_6$&$g_7$&$g_3$&$g_2$&$g_5$&$g_4$&$g_0$&$g_1$\\
\hline
$g_7$&$g_7$&$g_6$&$g_5$&$g_4$&$g_3$&$g_2$&$g_1$&$g_0$\\
\hline
\end{tabular}
\vspace{1ex}
\end{center}
There are seven independent one-forms in the space of ${\Omega^1}$ which we
denote by $\chi^g$ with subscripts $g_1,g_2,g_3,g_4,g_5,g_6,g_7$.
The components of the connection,
corresponding to the components of the Higgs, is denoted by two entries, one
for the entry of the space of $\Omega_1$ and the other for the
elements of group. According to the general formalism in section 2.2, we
can write the connection one forms as,
\be
A=\sum_{i=1}^7  \phi_{g_i} \chi^{g_i},
\ee
and $\Phi_{g_i}= 1 - \phi_{g_i}.$ The curvatures of $A$ are easily
obtained,
\be\ba{cl}
F_{g_ig_j}&=\Phi_{g_i} R_{g_i} \Phi_{g_j} -\Phi_{g_ig_j}, ~~if ~~
g_i\neq g_j\\[4mm]
F_{g_ig_j}&=\Phi_{g_i} R_{g_i} \Phi_{g_j} - 1,~~~if ~~
g_i = g_j.\ea
\ee
In order to get the potential for three Higgs doublets, we take the
following assumption,
$$\ba{cl}
\Phi_{g_1}(h)=\left(
\begin{array}{ccc}
0&{\Phi_1(x)}\\
\Phi^{\dag}_1(x)&0\ea\right),\ea $$
\be\ba{cl}
\Phi_{g_2}(h)=\left(
\begin{array}{ccc}
0&{\Phi_2(x)}\\
\Phi^{\dag}_2(x)&0\ea\right),\ea\ee
$$\ba{cl}
\Phi_{g_3}(h)=\left(
\begin{array}{ccc}
0&{\Phi_3(x)}\\
\Phi^{\dag}_3(x)&0\ea\right),\ea$$
$$\Phi_{g_4}(h)=\Phi_{g_5}(h)=\Phi_{g_6}(h)=\Phi_{g_7}(h)=0$$
where $h$ is any element of $G$. Then the only nontrivial curvatures are,
\be\ba{lll}
F_{g_1g_1}&=\Phi_{g_1}\cdot R_{g_1}\Phi_{g_1}-1,\hspace{5ex}
&F_{g_2g_2}=\Phi_{g_2}\cdot R_{g_2}\Phi_{g_2}-1,\\[3mm]
F_{g_3g_3}&=\Phi_{g_3}\cdot R_{g_3}\Phi_{g_3}-1,
&F_{g_1g_2}=\Phi_{g_1}\cdot R_{g_1}\Phi_{g_2},\\[3mm]
F_{g_1g_3}&=\Phi_{g_1}\cdot R_{g_1}\Phi_{g_3},
&F_{g_2g_1}=\Phi_{g_2}\cdot R_{g_2}\Phi_{g_1},\\[3mm]
F_{g_2g_3}&=\Phi_{g_2}\cdot R_{g_2}\Phi_{g_3},
&F_{g_3g_1}=\Phi_{g_3}\cdot R_{g_3}\Phi_{g_1},\\[3mm]
F_{g_3g_2}&=\Phi_{g_3}\cdot R_{g_3}\Phi_{g_2}.\ea\ee
If one use the most simple metric form on $\Omega^1$, i.e. $<\chi^g,
\chi^h>=E_g\delta^{gh^{-1}}$, where $E_{g_1}, E_{g_2}$ and $E_{g_3}$ are real
numbers and $\{E_{g_4}, E_{g_5}, E_{g_6}, E_{g_7}\}$ are zero, such that one
will get the perfect three Higgs potential,
\be \ba{cl}
V(\Phi_1,\Phi_2,\Phi_3)=&\alpha Tr{[E_1(\Phi_1\Phi_1^{\dag}-1)+
E_2(\Phi_2\Phi_2^{\dag}-1)+E_3(\Phi_3\Phi_3^{\dag}-1)]}^2\\[3mm]
&+\beta
[E_1^2Tr{(\Phi_1\Phi_1^{\dag}-1)}^2+E_2^2Tr{(\Phi_2\Phi_2^{\dag}-1)}^2\\
[3mm]
&+E_3^2Tr{(\Phi_3\Phi_3^{\dag}-1)}^2
+2E_1E_2Tr\Phi_1\Phi_2^{\dag}\Phi_2\Phi_1^{\dag}\\[3mm]
&+2E_1E_3Tr\Phi_1\Phi_3^{\dag}\Phi_3\Phi_1^{\dag}
+2E_3E_2Tr\Phi_3\Phi_2^{\dag}\Phi_2\Phi_3^{\dag}]\\[3mm]
&+\gamma [E_1^2Tr(\Phi_1\Phi_1^{\dag}-1)^2+E_2^2Tr(\Phi_2\Phi_2^{\dag}-1)^2+
E_3^2Tr{(\Phi_3\Phi_3^{\dag}-1)}^2 \\[3mm]
&+E_1E_2Tr(\Phi_1\Phi_2^{\dag}\Phi_1\Phi_2^{\dag}
+\Phi_2\Phi_1^{\dag}\Phi_2\Phi_1^{\dag})\\[3mm]
&+E_1E_3Tr(\Phi_1\Phi_3^{\dag}\Phi_1\Phi_3^{\dag}
+\Phi_3\Phi_1^{\dag}\Phi_3\Phi_1^{\dag})\\[3mm]
&+E_3E_2Tr(\Phi_3\Phi_2^{\dag}\Phi_3\Phi_2^{\dag}
+\Phi_2\Phi_3^{\dag}\Phi_2\Phi_3^{\dag})]\ea\ee
After taking the suitable value of $\{E_{g_1}, E_{g_2}, E_{g_3}, \alpha,
\beta$ and $\gamma\}$ one will get the three Higgs potential with nontrivial
vacuum expectation value.

\setcounter{equation}{0}
\section{ A Toy Model with Spontaneous $CP$ violation }
\hskip20pt
In order to get a model with spontaneous $CP$ violation,
the $SU(2)_L \times U(1)_Y$
electroweak gauge fields, quark-lepton spinors and three Higgs doublets
should be arranged as the fields over $M^4\times Z_2\times Z_2\times Z_2$
according to the three $Z_2$ symmetry in those fields.
Here for the simplicity we take the three $Z_2$ group as the same $Z_2$
symmetry, for example, the ${(CPT)}^2$ discrete symmetry in fermions.
Thus the fermions are arranged as,
\be\ba{cl}
\psi(x,g_0)=\psi(x,g_4)=\psi(x,g_5)&=\psi(x,g_6)=
\left(\ba{c}L\\R\ea\right)\\[5mm]
\psi(x,g_1)=\psi(x,g_2)=\psi(x,g_3)&=\psi(x,g_7)=-
\left(\ba{c}L\\R\ea\right)\ea\ee
and the gauge fields as,
\be
A_{\mu}(x,h)=\left(
\ba{cc}
L_{\mu}&0\\
0&R_{\mu}\ea\right), ~~ for~~~ h \in G\ee
where $L, R, L_{\mu}, R_{\mu}$ are
\be
L=\left( \ba{l}
U_i\\
D_i\\
N_i\\
E_i
\ea \right)_L \hs{3ex}
R=\left( \ba{l}
U_i\\
D_i\\
0\\
E_i
\ea \right)_R \hs{3ex} i=1,2,3
\ee
here
\be
U_i=\left( \ba{l}
u\\
c\\
t\\
\ea \right),\hs{2ex}
D_i=\left( \ba{l}
d\\
s\\
b\\
\ea \right),\hs{2ex}
N_i=\left( \ba{l}
{\nu}_e\\
{\nu}_{\mu}\\
{\nu}_{\tau}\\
\ea \right),\hs{2ex}
E_i=\left( \ba{l}
e\\
{\mu}\\
{\tau}\\
\ea \right).
\ee
and
\bea
L_{\mu}&=&-\f{ig}{2}I_2\ot {\tau}_iW^i_{\mu}\ot I_3-igB_{\mu}\left( \ba{ll}
\f{1}{6}&~\\
{}~&-\f{1}{2}\\
\ea \right) \ot I_2 \ot I_3;\nn \\
R_{\mu}&=&-ig_1B_{\mu}\left( \ba{lll}
\f{2}{3}&~&~\\
{}~&-\f{1}{3}&~\\
{}~&~&-1\\
\ea \right) \ot I_3.
\eea
 Furthermore, the Higgs fields $\Phi_n,~~ n=1, 2, 3$ ~are,
\be
\Phi_n(x)=\left\{\left(\ba{ccc}
\left(\begin{array}{ccc}
{\phi}_n^{0*}&{\phi}_n^{+}\\[1mm]
-{\phi}_n^{+*}&{\phi}_n^{0}\end{array}\right)\\
&\left(\ba{c}
\phi_n^+\\
\phi_n^0\ea\right)\ea\right)\otimes I_3^G\right\}
\cdot\left(\begin{array}{ccccc}
\Gamma_n^{U}\\[1mm]
&\Gamma_n^{D}\\[1mm]
&&\Gamma_n^{L}\end{array}\right).\ee
where $\Gamma_n^{U}, \Gamma_n^{D}, \Gamma_n^{L}$ are the coefficient of
Yukawa couplings.

After some algebraic calculations one gets the Lagrangian for that
model. The Yang-Mills terms for $SU(2)_L \times U(1)_Y$ gauge fields,
the kinetic terms for fermions and its coupling to gauge fields remain
the same as
in \cite{Ding}, i.e.,
\be
 L_{YM}(x)=\displaystyle-\frac{1}{4N_L}3g^{2}W^{i}_{\mu\nu}W^{i\mu\nu}
-\frac{1}{4N_Y}\frac{19g'^{2}}{3}B_{\mu\nu}B^{\mu\nu}\ee
and
\be
 L_{Fermion}(x)=\overline L i \gamma^{\mu}(\partial_{\mu}
+L_{\mu})L + \overline R i\gamma^{\mu}
(\partial_{\mu}+ R_{\mu})R \ee
We are going to give other terms, which are not the same as in standard
model. First, the Yukawa coupling terms are,
\be
\ba{ll}
L_{Yukawa}&=\displaystyle\sum_{n=1}^3 \left( \ba{ll}
{\bar{\psi}}_L&{\bar{\psi}}_R \ea \right)
\left( \ba{ll}
{}~&{\P}_n\\
{\P}^{\da}_n&~
\ea \right)
\left( \ba{l}
{\psi}_L\\
{\psi}_R
\ea \right) \\[4mm]
&=\displaystyle\sum_{n=1}^3 {\bar{\psi}}_L{\P}_l{\psi}_R+
{\bar{\psi}}_R{\P}^{\da}_l{\psi}_L\\[3mm]
&=\displaystyle{\G}^U_{nij}({\bar{U}}^i_L{\p}^{0\ast}_n-
{\bar{D}}^i_L{\p}^{\da \ast}_n)
U^j_R+{\G}^D_{nij}({\bar{U}}^i_L{\p}^{\da}_n+{\bar{D}}^i_L{\p}^0_n)D^j_R
\\[3mm]
{}~&\displaystyle +{\G}^L_{nij}({\bar{N}}^i_L{\p}^{\da}_n
+{\bar{E}}^i_L{\p}^0_n)E^j_L
+h.c.,
\ea
\ee

In general the Yukawa coupling in (3.9) can not preserve the neutral flavor
current conservation (NFC) provided the matrices ${\G}^F_n,~ (F=U,D,L)$
can not be diagnolized simultaneously by a biunitary or
biothogonal transformation. So Weinberg and Glashow introduced a discrete
symmetry in Lagrangian to ensure the NFC. But there is another way to
overcome this difficulty by imposing some constraints on Yukawa
coupling matrix
so that the matrix ${\G}$ can be diagnolized simultaneously\cite{Wu,Wu2}.
In other words we require that the real matrices ${\G}^a_F$ be written
into the following structure
\be
{\G}^F_n= {\sum}^3_{\a=1}g^F_{l\a}O^F_L{\o}^{\a}
(O^F_R)^T,\hs{3ex} n=1,2,3. \ll{pp}
\ee
with ${{\o}^{\a},\a=1,2,3}$ the set of diagnolized projection matrices
${\o}^{\a}_{ij}={\d}_{i\a}{\d}_{j\a}$, $O^F_{L,R}$ are the arbitrary
orthogonal matrices and independent of the Higgs doublet label $n$.
This is the crucial point for ensuring the NFC.

For the kinetic terms of Higgs fields we get
\be
L_H(x)=\frac 1 N \sum _{n=1}^3
2E_n\sigma_n(D_{\mu}\pi_n)^{\dag}D^{\mu}\pi_n
\ee
where we introduce an usually notation of doublet scalar field $\pi_n$,
\be
\pi_n=\left(\begin{array}{cl}
{\phi}_n^+\\
{\phi}_n^0\end{array}\right),\ee
$$D_{\mu}\pi_n=({\partial}_{\mu}-\frac{ig}{2}{\tau}_{i}W_{\mu}^{i}
-\frac{ig'}{2}B_{\mu})\pi_n,$$
and the notations $\sigma_n$
\be
\sigma_n=Tr\left(\begin{array}{ccccccc}
\Gamma_n^U{\Gamma_n^U }^{\dag}&\\[1mm]
&&\Gamma_n^D{\Gamma_n^D }^{\dag}&\\[1mm]
&&&&\Gamma_n^L{\Gamma_n^L}^{\dag}\end{array}\right),\ee
If the ${\G}^F_n,~ (F=U,D,L)$ have been chosen, we should take the
values of $E_1, E_2, E_3$ such that,
\be
 \frac 2 N E_n\sigma_n=1 \ee
for all
$n=1, 2, 3$ in order to get the normalized kinetic terms for the three
Higgs fields.

Finally, from (2.25) and above assignments,
we get the potential of three Higgs fields $\pi_n$,
\be \ba{cl}
V(\pi_1,\pi_2,\pi_3)&=\displaystyle\sum_{n=1}^3[a_{nn}Tr(\pi_n\pi_n^{\dag})^2-
a_nTr(\pi_n\pi_n^{\dag})]\\[3mm]
&+\displaystyle\sum_{n<m}\{a_{nm}Tr(\pi_n\pi_n^{\dag})(\pi_m\pi_m^{\dag})+
b_{nm}Tr(\pi_n\pi_m^{\dag})(\pi_m\pi_n^{\dag})\\[3mm]
&+[c_{nm}Tr\pi_n\pi_m^{\dag}\pi_n\pi_m^{\dag}+h.c.]\}\ea\ee
where
\bea
a_{nn}&=&2(\alpha+\beta+\gamma)E_n^2p_{nn} \nn\\
a_{n}&=&4(\alpha+\beta+\gamma)E_n^2\sigma_n+4\alpha
\displaystyle\sum_{m\neq n}E_mE_n
\sigma_n \nn\\
a_{mn}&=&4\alpha E_mE_n p_{mn} \\
b_{mn}&=&4\beta E_mE_n q_{mn}  \nn\\
c_{mn}&=&2\gamma E_mE_n s_{mn}   \nn
\eea
and
$$
p_{mn}=Tr(({\G}^U_m{\G}^{U\da}_m)({\G}^U_n{\G}^{U\da}_n)+
({\G}^D_m{\G}^{D\da}_m)({\G}^D_n{\G}^{D\da}_n)
+({\G}^L_m{\G}^{L\da}_m)({\G}^L_n{\G}^{L\da}_n))$$
$$q_{mn}=Tr(({\G}^U_m{\G}^{U\da}_n)({\G}^U_n{\G}^{U\da}_m)+
({\G}^D_m{\G}^{D\da}_n)({\G}^D_n{\G}^{D\da}_m)
+({\G}^L_m{\G}^{L\da}_n)({\G}^L_n{\G}^{L\da}_m))$$
$$s_{mn}=Tr(({\G}^U_m{\G}^{U\da}_n)({\G}^U_m{\G}^{U\da}_n)+
({\G}^D_m{\G}^{D\da}_n)({\G}^D_m{\G}^{D\da}_n)
+({\G}^L_m{\G}^{L\da}_n)({\G}^L_m{\G}^{L\da}_n))$$
We find that the terms propotional to $c_{mn}$ possess
CP nonconservation\cite{Wein}. Futhermore we read that the potential is
invariant under some discrete symmetry transformation\cite{Wein}.

\section{Remark and Conclusion}

In this paper, we construct a three-Higgs
toy model of SCPV with the help of the differential
calculus on discrete group in the spirit of non-commutative geometry.
In our model the NFC is ensured by imposing some constraints on
the Yucawa coupling matrices (just as in \cite{Wu}) and we allow each
Higgs doublet couple to all the fermions.  By a straightforward
calculation we obtain a potential which has a $CP$ violation term and
is invariant
under a discrete symmetry transformation. However we should address
that the NFC
is not necessary in our discussions. If we abandon the NFC, i.e. abandon
the so called simultaneously diagnolized condition of Yukawa coupling
matrix, we can obtain the Weinberg-Hall model\cite{weh}.

Futhermore the Yukawa coupling will be carefully choosed
so that the following two physical conditions satify:\\
1. if $CP$ violate, the values of $c_{mn}$ should satisfy the triangular
condition\cite{Wein}.\\
2. the masses of three Higgs bosons cannot differ too much, therefore the
expectation values of Higgs doublets must be at the same level.\\
Then we may construct a physical $CP$ violation model. This question will be
investigate more.

\end{document}